\documentclass[10pt,journal]{IEEEtran}
\usepackage{array}
\usepackage{graphicx}
\usepackage{epsfig}
\usepackage{cite}
\usepackage{epstopdf}
\usepackage{amsfonts,amsmath,amsthm,amssymb,balance}
\usepackage{multirow,balance}
\usepackage{bm}
\usepackage[usenames,dvipsnames]{color}
\usepackage{colortbl}
\usepackage{float}
\usepackage{subfigure}
\usepackage{tabularx}
\usepackage{enumerate}
\usepackage{algorithm}
\usepackage{algorithmic}
\usepackage[marginal]{footmisc}
\usepackage{url}

\usepackage[colorlinks,citecolor=magenta,linkcolor=blue]{hyperref} 

\graphicspath{{figures/}}
\def\BibTeX{{\rm B\kern-.05em{\sc i\kern-.025em b}\kern-.08em
		T\kern-.1667em\lower.7ex\hbox{E}\kern-.125emX}}

\begin{document}
	\title{Enhancing LEO Mega-Constellations with Inter-Satellite Links: Vision and Challenges}
	
		\author{ Chenyu Wu, \IEEEmembership{Member, IEEE}, Shuai Han, \IEEEmembership{Senior Member,~IEEE}, Qian Chen, \IEEEmembership{Member, IEEE}, Yu Wang, 
	\\	Weixiao Meng, \IEEEmembership{Senior Member,~IEEE}, and Abderrahim Benslimane, \IEEEmembership{Senior Member,~IEEE} 
		
		\thanks{
			
			This work of C. Wu was supported by the Postdoctoral Fellowship Program of CPSF under grant number GZC20233483. 
			\textit{(Corresponding author: Shuai Han.)}
			
			C. Wu, S. Han, Y. Wang and W. Meng are with the School of Electronics and Information Engineering, Harbin Institute of Technology, Harbin 150001, China (e-mail: wuchenyu@hit.edu.cn; hanshuai@hit.edu.cn; wangyu\_hitcrc\_@stu.hit.edu.cn; wxmeng@hit.edu.cn). 
			
			Q. Chen is with the Department of Electrical and Electronic Engineering, The University of Hong Kong, Hong Kong (email: qchen@eee.hku.hk). 
			
			A. Benslimane is with University of Avignon, France. (email: abderrahim.benslimane@univ-avignon.fr).

		}
	}
	
	\markboth{} {Shell \MakeLowercase{\textit{et al.}}: Bare Demo of IEEEtran.cls for IEEE Journals}
	
	\renewcommand{\baselinestretch}{1.2}
	
	\maketitle
 	
	\begin{abstract}
	Low Earth orbit (LEO) satellites have been envisioned as a significant component of the sixth generation (6G) network architecture for achieving ubiquitous coverage and seamless access. However, the implementation of LEO satellites is largely restricted by the deployment of ground stations. Inter-satellite links (ISLs) have been regarded as a promising technique to fully exploit the potentials of LEO mega constellations by concatenating multiple satellites to constitute an autonomous space network. In this article, we present the merits of implementing ISLs in LEO mega constellations and the representative applications empowered/inspired by ISLs. Moreover, we outline several key technical challenges as well as potential solutions related to LEO satellite networks with ISLs, including performance analysis for system design, routing and load balancing, and resource allocation. Particularly, the potential of using ISLs in enhancing in-flight connectivity is showcased with a preliminary performance evaluation. Finally, some open issues are discussed to inspire future research.
	\end{abstract}
	
	\IEEEpeerreviewmaketitle
	
	\section{Introduction}\label{sec:intro}
	 Satellite network, as a complement of the terrestrial cellular network, plays a pivotal role in realizing the six-generation (6G) prospects, such as ubiquitous coverage and broadband access\cite{sagin_survey}. Compared with geosynchronous Earth
	 orbit (GEO) or medium Earth orbit (MEO) satellites, low Earth orbit (LEO) satellites admit less path loss, lower latency, and higher data rate due to their lower height, and hence have drawn great interest from both industry and academic research\cite{dense}. However, to embrace seamless global communication coverage, it is indispensable to deploy hundreds of LEO satellites to constitute mega constellations. In recent years, the construction of LEO mega constellations has upsurged, promoted by some leading companies in the industry field. For example, SpaceX has begun the Starlink project and expects to launch 42,000 Starlink satellites by mid-2027. OneWeb announced to build a constellation comprising 648 LEO satellites to provide carrier-grade service to global users.

	In general, the quantity of ground stations (GSs) is significantly fewer than that of satellites since the deployment of satellite GSs is constrained by excessive investments and many geographical factors. In this case, the limited feeder links become the bottleneck for making full use of the LEO mega-constellations. For example, the telemetry data of an LEO satellite may not be fully collected by GSs within the short contact time and thus lose its vitality. In this context, inter-satellite links (ISLs) are envisioned to make full use of the advantages of LEO mega constellations by interconnecting hundreds of satellites to forge a space information network. Employing ISLs, GSs are able to download data from and send commands to the satellites that are out of sight, which enables real-time tracking of the whole satellite network. Therefore, ISLs have been envisioned as a viable paradigm to promote the advance of the next-generation satellite network. In a nutshell, the salient features of ISLs in LEO mega constellations can be summarized as follows.
	\begin{itemize}
		\item \textbf{\underline{Network autonomy:}} By embedding on-board processing units as payload, satellites can form a decentralized network structure to forward user data directly via ISLs such that the data transmission can bypass the GSs. This reduces the satellite network's dependency on the ground segment and also the number of gateways required. 
		
		\item \textbf{\underline{Enhanced quality of service:}} Thanks to the flexibility and high capacity of ISLs, data can be disseminated from one satellite to another with an ultra-low latency through parallel transmission.
		As such, ISLs enable real-time data sharing among satellites, thereby releasing the burden of a single feeder/backhaul link and enhancing the quality of service (QoS). Moreover, with ISLs, satellite networks can well accommodate non-uniform traffic distributions through efficient offloading and distributed processing.
		\item \textbf{\underline{Favorable propagation environment:}} Wireless signals propagating in the vacuum are not affected by the rain and tropospheric attenuation. Besides, directly transmitting by ISLs rather than being relayed by
        ground-to-satellite links benefits from a shorter propagation distance and enhanced sustainability.
		
	\end{itemize}


\begin{table}[ht]
	\caption{Comparison between LISLs and RF-based ISLs} \label{tab:compare}
	\begin{center}
		{\footnotesize	\begin{tabular}{|c|c|c|}\hline  
				\textbf{Characteristic} & \textbf{LISL} & \textbf{RF-based} \\ \hline
				Band &  Terahertz (THz) & Ka, millimeter-wave \\ \hline
				Spectrum & Unlicensed & Licensed \\ \hline
				Data Rate & Faster (Gigabit-per-second) & Lower \\ \hline
				Antenna size & Small & Big\\ \hline
				Directivity & High beam concentration & Beam spread \\ \hline
				Energy efficiency & High & Low \\ \hline
				Security & High & Low  \\ \hline
				Interference & No & Severe \\ \hline
				Solar radiation & Vulnerable & No impact \\ \hline
				Link setup time & Long & Short \\ \hline
		\end{tabular}}
	\end{center}
\end{table}

For decades, microwave transmission based on radio frequency (RF) technology has been the primary scheme for ISLs. With the recent progress in fiber optics technologies, laser ISLs (LISLs) are gradually replacing RF links to become the dominant form\cite{lisl_mag}. Due to the high frequency and large bandwidth, LISLs have many advantages over RF-based ISLs, as compared in Table \ref{tab:compare}.   LISLs are anticipated to support capacities of up to 10 Gigabit-per-second, with expectations to gradually expand to Terabit-per-second. Based on the relative positions of satellites within a constellation, ISLs can be classified into intra-orbit ISLs and inter-orbit ISLs, as illustrated in Fig. \ref{application}. Since the relative distance between two satellites in the same orbit is approximately fixed, intra-orbit ISLs are expected to be more stable than inter-orbit ones\cite{broadband}. Emerging projects of satellite constellations have already planned to support ISLs. 
For instance, each Starlink satellite and Telesat Lightspeed satellite would be equipped with terminals that support four LISLs. There is no doubt that ISLs will usher in a new era of satellite Internet services for LEO mega constellations.
	
Motivated by the above, this paper aims to comprehensively review the visions and challenges of employing ISLs in LEO constellations. We first highlight the unique features and representative application scenarios of ISLs. Then, we discuss new considerations and challenges pertaining to system performance analysis, routing, and resource allocation. Particularly, numerical results are presented to show the performance gain brought by ISLs. Finally, several open issues are listed for future study.

	\section{Applications of ISLs}

	\begin{figure*}[t]
	\centering
	\includegraphics[width=1\textwidth]{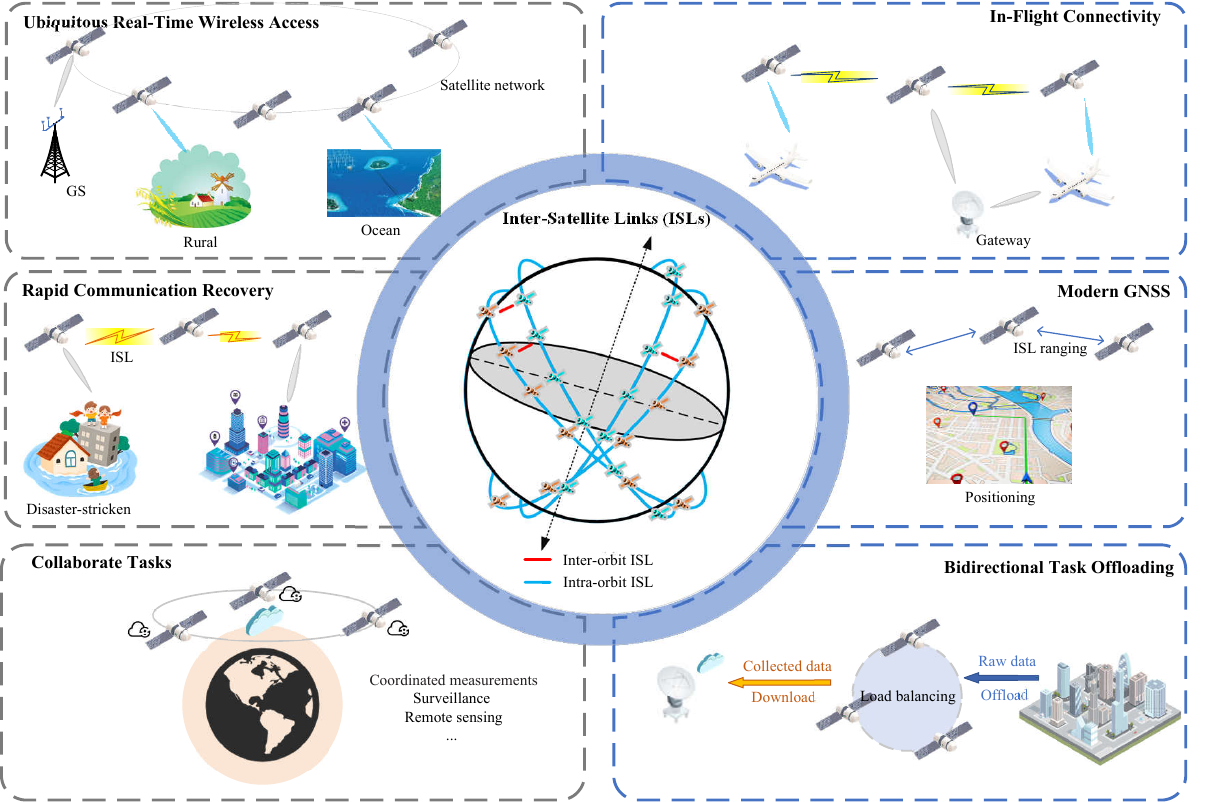}\\
	\caption{Typical application scenarios for ISLs.}\label{application}
\end{figure*}	

Due to the aforementioned advantages, ISLs have a wide range of applications in satellite networks, as shown in Fig. \ref{application}, which can be categorized into: 1) ISL-empowered applications. These  can hardly be realized by LEO satellites without ISLs; and 2) ISL-inspired applications, for which existing operations or systems can be promoted by ISLs. Examples and detailed introductions are given as follows.

	\subsection{ISL-Empowered Applications}

\subsubsection{Ubiquitous Real-Time Wireless Access}

Satellite networks have been envisioned to realize ubiquitous coverage, which means that user equipment can enjoy real-time wireless access at any time and any place. However, this hinges on the dense distribution of LEO satellites, and more importantly, their serving ground stations. In those extreme regions, where the deployment of GSs is restricted, the feeder link is likely to be unavailable, and consequently, real-time wireless access is impossible without ISLs. Under such cases, the nearest satellite can connect to a satellite that has access to a gateway via ISLs to establish an available feeder link, hence enabling real-time data transmission. 


\subsubsection{Rapid Communication Recovery} 

In times of emergency (e.g., when disaster strikes), the terrestrial infrastructures are either destroyed or overloaded. Traditional emergency communication mainly relies on complementary components such as high/low altitude platforms, unmanned aerial vehicles, and customized handheld devices. The discovery of catastrophes and the deployment of those external will cause a serious delay, and inevitably affect rescue operations and disaster relief. With ISLs, emergency communication can be rapidly recovered with no need for human intervention. Hence, victims can proactively report the status of the disaster and their instantaneous positions, which helps accelerate the rescue process.

\subsubsection{Collaborate Tasks for Satellites} Missions that involve multiple satellites working in tandem, such as those studying the Earth’s magnetosphere or conducting astrophysical observations, may need to exchange data or coordinate measurements, which can not be fulfilled without ISLs. ISLs enable real-time data sharing and task coordination between satellites, hence providing satellite networks with  improved spatial and temporal resolutions of surveillance and remote sensing.

	\subsection{ISL-Inspired Applications}
\subsubsection{Modern GNSS} 
Modern Global Navigation Satellite System (GNSS) has featured ISL to boost the positioning performance and foster the autonomy of the space segment. GNSS relies on accurate clock synchronization to achieve successful positioning, which is generally managed by a ground network of monitoring stations. Owing to the free-space and high-capacity transmission capability, ISL enables sub-millimeter ranging between satellites, hence enhancing time synchronization and orbit determination of the space segment. Moreover, whenever the GPS signals are weak, LEO satellites can provide an alternate way to determine the location of mobile devices by using the Doppler measurements.


\subsubsection{In-Flight Connectivity} 

ISL can help boost some commercial applications which can not be fully achieved by 
ground segment, e.g., in-flight connectivity (IFC) and on-board connectivity. Due to the high costs of aircraft gateways, employing space-to-air communications is more cost-effective for IFC. However, the connected satellites en-route may not always be visible to a GS/gateway, which necessitates the utilization of ISLs. Moreover, by invoking the caching capabilities of the satellite network and utilizing multiple ISLs for file downloading, the transmission delay of the requested file can be significantly decreased.

\subsubsection{Bidirectional Task Offloading}  

ISLs can promote various offloading-enabled applications, e.g., edge computing, federated learning, etc. The disparity between the size of a task and the length of a satellite's contact time with the gateways constitutes a main bottleneck for bidirectional task offloading, which can be well compensated by using ISLs. On the one hand, ISL can allow satellites to collaboratively and adaptively offload their collected data (e.g., target surveillance, environmental monitoring) and computation-intensive missions (image processing) to GSs according to their contact time with the Earth station. On the other hand, missions that can hardly be accomplished by the ground segment can be offloaded to satellites. Particularly, computing tasks offloaded to space can be diverted from congested satellites to idle ones via ISLs so that the network resources in space can be fully utilized.

	\section{New Considerations and Challenges}

In this section, we present main design challenges when considering ISLs in satellite networks. Promising solutions for tackling these challenges are also proposed to inspire future work.

\subsection{Performance Analysis for System Design}

Performance analysis tailored to satellite networks with ISLs is primarily required to gain useful insights for achieving low-latency routing and efficient resource utilization.

\subsubsection{Analysis of ISL Paths}

Any two ground users, even though they are far apart (e.g., London and Singapore), are able to be connected through multi-hop ISL relays. Although the data rate of ISLs is high, the propagation latency will be unaffordable when the routing paths experience long distances. Therefore, the shortest distance path and the minimum hop path (MHP) become important criteria for measuring the propagation delay and transmission efficiency. Different from previous studies that rely on complex network simulations, the authors in \cite{9351765} proposed a theoretical analysis approach to estimate the ISL hop-count between ground users in an inclined orbital constellation. Taking Starlink phase I-B constellation as an example, the determination of hop-count and the effects of constellation parameters were investigated. Particularly, simulation results showed that the difference in the paths from different starting satellites can be up to 45 hops. Such a substantial disparity in the number of hops further underscores the importance of designing appropriate association strategies for ground users. The authors further presented a discriminant function to judge whether the SDP belongs to the MHP in a given satellite constellation network \cite{9681176}. Simulations verify that for low-inclination (less than 68 degrees) or large phasing offsets satellite constellations, all SDPs belong to the set of MHP. With this conclusion, the calculation of the minimum hop count can be simplified.  These works offered guidelines for the routing design in satellite networks through theoretical analysis of ISL paths. Although SDP and MHP can preliminarily reflect the transmission latency, the link quality and sustainable time are not taken into account in most existing works. As introduced in Section \ref{sec:intro}, intra-orbit ISLs are more stable than inter-orbit ISLs during task transmission. Thus, to achieve efficiency without compromising much reliability, a routing mode with enough link stabilization and a small ISL hop count is worth investigating, which will be elaborated in the next subsection. Note that different from other terrestrial or aerial networks with dynamic topology, the movements of LEO satellites are ruled by deterministic orbits, which offers great opportunities for deriving analytic approaches to characterizing the performance of LEO constellations, rather than through extensive simulations. In this regard, other important criteria such as coverage range, capacity, bit error rate, and outage probability, can be analyzed by exploiting the topology regularity of LEO mega constellations.


\subsubsection{Gateway Placement}

Ground infrastructures such as gateways are necessarily required by the satellite network to get access to the Internet backbone. On the one hand, dense gateways should be reasonably deployed to provide better Internet access, especially in places with huge populations. On the other hand, the deployment of ground stations generates high costs and is highly restricted by geographical factors. Therefore, gateway placement constitutes an important branch of constellation design. The authors in \cite{gateway} proposed a gateway placement optimization method to strike a balance between the number of gateways and the traffic loads, but ISLs are not considered. As ISLs make the decentralized satellite networks possible, ISLs can effectively reduce the number of ground gateways required. Conversely, deploying more gateways can release the usage of ISL transmission. Therefore, a comprehensive model is essential to incorporate the costs of deploying a satellite gateway and operating ISLs from a long-term perspective. Besides, analytical frameworks are required to determine the optimal quantity of satellite gateways and their sites while considering several practical constraints, such as link interference, satellite bandwidth, and the number of ISLs.

\subsubsection{Constellation Design} ISLs are often ignored in traditional constellation design. In general, polar orbit planes are preferred compared with inclining orbit planes to avoid intermittent ISLs, but are not desirable for LEO satellites\cite{constellation_design}. For constellation design, the inclination and angular spacing of each orbit plane should be taken care of to realize global coverage with a minimum number of satellites as well as to avoid ISL congestion at specific regions.


\begin{figure*}[t!]
	\centering
	\subfigure[Traditional routing scenario, where satellites form a grid-like topology by establishing two inter-orbit ISLs and two intro-orbit ISLs. ISL breakdowns may result in ISL handover and seam areas.]{\label{route_static}
	\includegraphics[width = 0.48\textwidth]{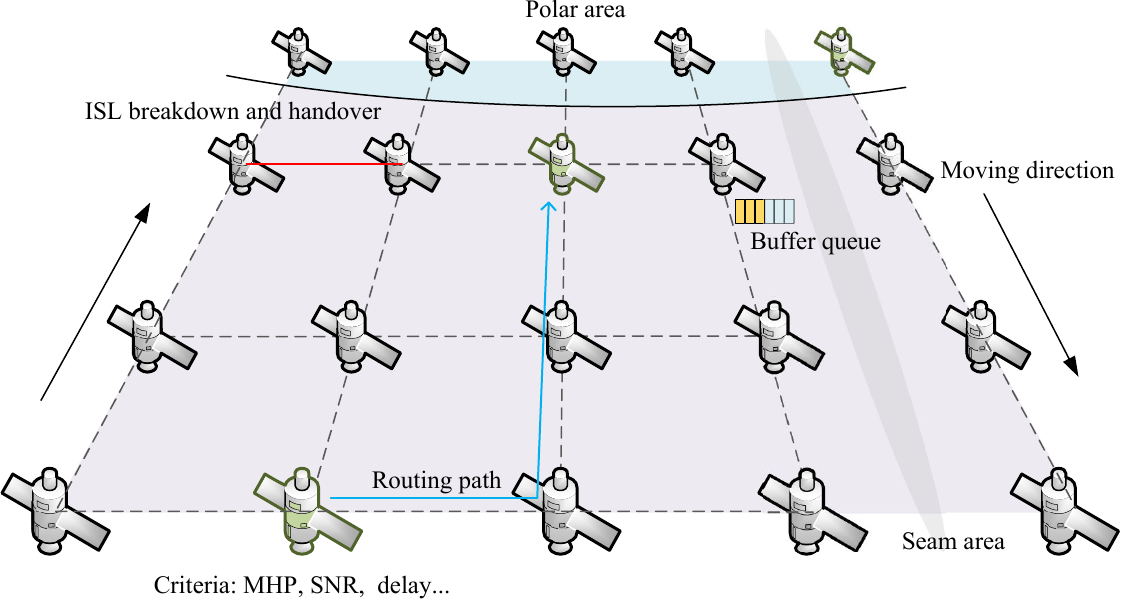}}
	\subfigure[A satellite-terrestrial cooperative routing framework, where MEO/GEO satellites act as the routing manager and GSs assist the cooperative routing.]{\label{route_dynamic}
	\includegraphics[width = 0.47\textwidth]{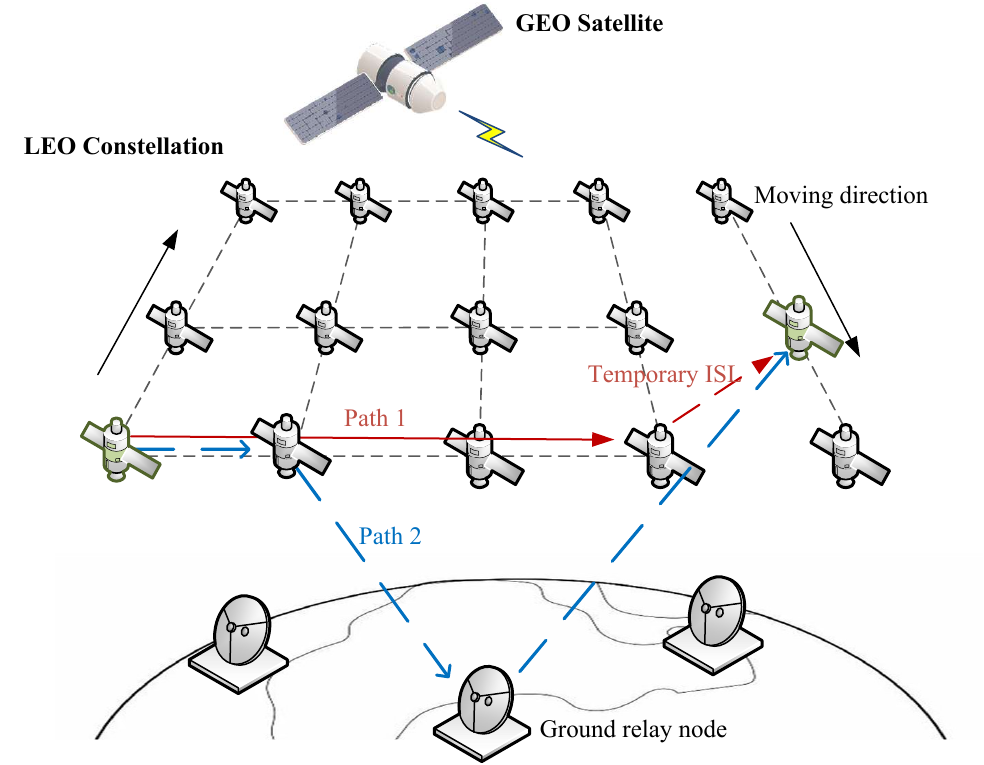}}
	\caption{Illustration of a routing scenario in LEO mega constellations}\label{fig:routing}
\end{figure*}
	
\subsection{Routing and Load Balancing}
ISLs raise new issues of ISL handover and routing design in LEO constellations, as illustrated in Fig. \ref{fig:routing}. Handover deals with the movement and reestablishment of ISLs due to the variation of visibility between satellites, while routing addresses the problem of finding an effective path for dependable end-to-end transmissions. These two steps collaborate to avoid service interruption and improve the QoS.  For LEO satellite integrated networks, the dynamic topology and heterogeneous distribution of user QoS/ satellite processing capabilities make routing a difficult task. 

\subsubsection{From Static Routing to Dynamic and Hybrid Routing} Different from terrestrial networks, the trajectories of satellites and the time-varying network topology of the satellite network are predictable. This distinct feature can be leveraged for facilitating pre-scheduled handover and routing design. For example,
when two satellites move in the opposite direction or move across the polar region, the inter-orbit ISL is vulnerable and is more likely to experience handover. Routing schemes can be categorized into \textit{static routing}, which utilizes the pre-calculated topology and \textit{dynamic routing}, which utilizes instantaneous state information. 

For static routing, the snapshot-based idea is widely adopted, for which the orbit period is cut into several time slots, and within each time slot, the network topology is considered to be stable. Then, for each snapshot, traditional routing algorithms such as the Dijkstra algorithm and heuristic algorithms like generic and ant colony algorithms can be employed. 
However, these algorithms simply assume that ISLs follow on/off switches and ignore the ISLs' attributes. The graph model and tree structure can intuitively characterize the distribution of link nodes and the link quality, and hence have been explored for routing design. In \cite{10121666}, the authors orchestrated these two ideas by adopting a coordinate graph (CG) model to characterize the topology, based on which a binary-tree is constructed to determine the minimum-hop paths.  Nevertheless, static routing can not adapt to the real-time network situation and handle burst disconnection, which calls for dynamic/online routing strategies. For dynamic routing schemes, 
the network states, such as available link ports and payload of all satellites, are periodically collected and flooded to each node to maintain an up-to-date routing table. However, flooding among the whole network generally results in excessive convergence time and huge overhead. To overcome this issue, a multi-layer satellite structure and hybrid routing mechanism can be invoked, where MEO/GEO satellites act as the manager for routing decisions and GSs act as potential relay nodes for cooperative routing, as demonstrated in Fig. \ref{route_dynamic}. Besides, considering that traditional routing algorithms suffer from high complexity and long convergence time, machine learning is envisioned to be an effective tool to predict the variation of network topology and realize routing with fast convergence and robustness\cite{ai}. 

\subsubsection{Load Balancing and QoS Provision}

Several other practical considerations need to be taken into account for routing design in LEO constellations. First, the distribution and demands of users are unevenly distributed, resulting in imbalanced regional traffic loads. As such, the paths with the shortest distance or hop-counts found by traditional routing algorithms may be congested, which calls for new routing strategies for load balancing. Second, there exist different traffic classes in satellite-air integrated networks such as delay-sensitive and throughput-hungry. Then, the routing design should also account for various metrics such as delay, delay jitter, capacity, hop counts, and queuing delay. Therefore, how to distribute the traffic flows among ISLs and space-to-ground links according to their different QoS requirements is well worth studying. 


	\subsection{Resource Allocation}
	
Satellite network possesses resources from dimensions of time, frequency, and power, as well as the new spatial degree of freedom of link scheduling provided by ISLs. Despite the high data rate of ISLs, the resources for a satellite network are still scarce and require deep dug to improve the spectrum efficiency and energy efficiency.

\subsubsection{Link Scheduling} A typical pattern for establishing ISLs (e.g., in the Iridium system) is the grid-like pattern, where each satellite can maintain four quasi-permanent ISLs (two intra-plane ones and two inter-plane ones), as shown in Fig. \ref{route_static}.  Although this regular architecture possesses enhanced stability, the flexibility of system design as well as network performance is largely limited. With the development of advanced on-board transceivers, dynamic ISL architecture is foreseen to be realized, where a satellite can dynamically establish a temporary ISL with any satellites within its communication range. Compared with permanent and single-path ISLs, dynamic and multi-path ISLs will play a vital role in achieving low-latency content delivery. In this case, the most important factor when designing satellite networks with dynamic ISLs is how to optimize the \textit{link assignment/satellite association} in a cost-effective manner, which faces great challenges. First, the link assignment introduces large-scale binary variables, especially for a mega constellation with hundreds of satellites. Meanwhile, the number of ISLs that a satellite can establish is generally limited for reasons of investments and implementation, which introduces extra constraints. Relaxation or replacement-based approaches such as the penalty-based method can be invoked to deal with this type of binary programming. Some discrete approaches, e.g., the Hungarian algorithm and many-to-many matching can also be used when considering  two satellite groups with one-hop transmission. Second, as power supply for satellites is usually guaranteed by solar panels, many practical issues should be considered for ISL link assignment to ensure stability and energy efficiency, such as link distance, satellites' angular velocity, and link setup time. For instance, the link distance not only affects the link quality (link interruption probability) but also the power consumption of the communication device. As such, there exists a fundamental trade-off between the transmission distance (or equivalently, communication performance) and operation cost  of an ISL. Moreover, the inter-orbit links should be less selected compared with the intra-orbit ISLs to avoid intermittent connections.
Finally, a large number of ISLs also pose severe interference in space, especially for the RF bands. To tackle this, dedicated multiplexing schemes for laser beams as well as efficient beam management/steering strategies should also be devised.

\subsubsection{Task Offloading/Downloading}  As mentioned above, ISLs facilitate many up-to-date applications of the satellite network, such as modern GNSS, remote sensing, edge computing, etc. In such cases, the link scheduling scheme and resource allocation require joint optimization. For example, the authors of \cite{7805169} proposed a collaborative data downloading scheme, where data offload among the satellites via ISLs and data download to the earth station were jointly scheduled. Note that the resource allocation associated variables are generally coupled with link assignment, which aggravates the difficulty of solving the corresponding optimization problem. The exhaustive search-based methods such as branch and bound are computationally prohibitive due to the scale of the satellite network. To reduce the design complexity, clustering-based methods can be employed to form distributed satellite networks. However, this generally leads to sub-optimal solutions. Moreover, essential analysis regarding ISLs can be conducted first to transform the original problem into a more tractable form.  Besides huge complexity, several constraints tailored to specific applications of ISLs should be considered. For example, for enhancing the GNSS, new ranging constraints (i.e., switching ISLs to sufficient different satellites within a short time) and flow conservation constraints should be considered to ensure precise orbit determination and maintain flow stability\cite{9211777}. In addition, ISLs provide new opportunities for distributed task processing since ISLs enables flexible sharing of computing resources according to the mission requirements. However, for computation-oriented tasks, the deployment of computing units and the offloading strategies for a specific task must be designed with precision.

\subsubsection{A Case Study}

\begin{figure}[t]
	\centering
	\includegraphics[width = 0.5\textwidth]{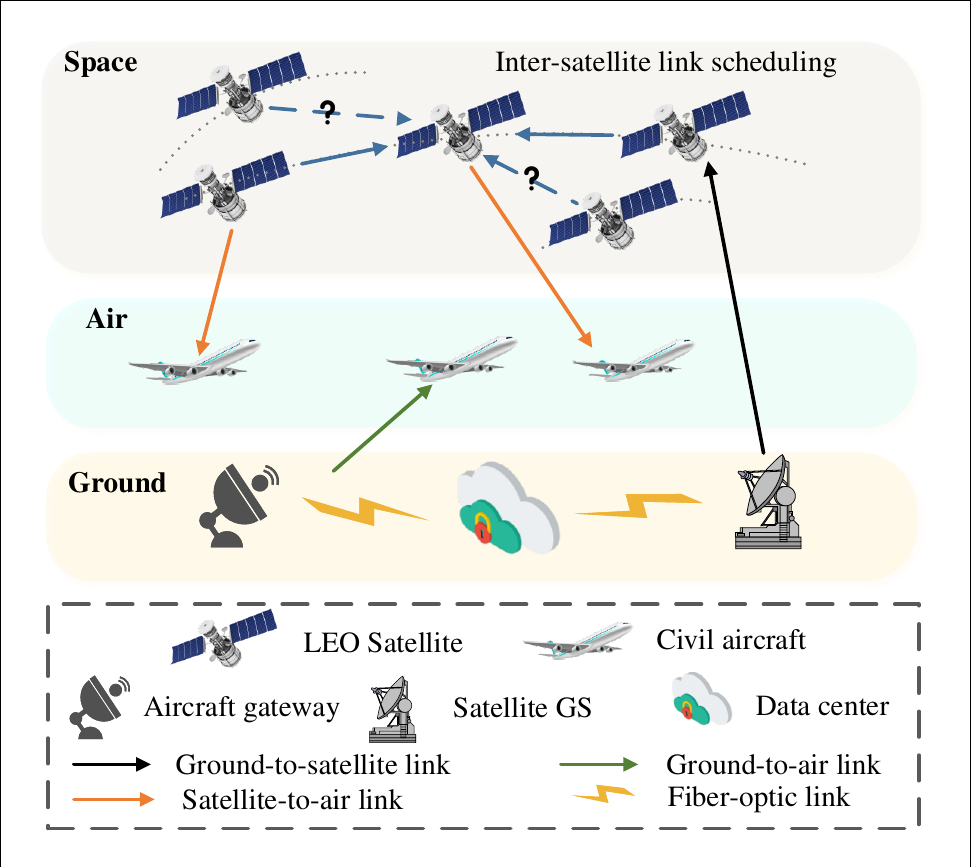}
	\caption{Proposed framework to realize IFC by exploiting ISLs. \label{fig:SAGIN_Architecture}} 
\end{figure}

As illustrated in Fig. \ref{fig:SAGIN_Architecture}, we consider a framework for achieving uninterrupted IFC, where the space and ground networks cooperatively deliver content desired by the passengers in an aircraft. Based on the content type, the files requested are divided into cached contents and non-cached contents. Specifically, cached files are instantaneously accessible via satellites' storage, while non-cached files can only be retrieved from GSs. The detailed transmission schemes for these two types of files are given in \cite{qian}. ISLs are implemented in the transmission process for sharing the contents among the satellite network. We formulate an optimization problem of the link assignment (i.e., satellite association by ISLs) and file downloading ratio for each type of file to minimize the average delay. For non-cached files, an extra bandwidth allocation factor is considered. To solve this problem, the property of the optimal transmission delay was first analyzed, which helps eliminate the variables of the downloading ratio and thus reduces the scale of the optimization problem. 

For the simulation setups, a total of 120 satellites are distributed across 6 orbital planes, with each plane containing 20 satellites at an altitude of 1000 km, and the orbit inclination is set at 53$^{\circ}$. We consider the Airbus A320 aircraft, each of which can generate at most one file request in a time slot. The number of packets for each file follows a uniform distribution within the ranges of [50, 100], [500, 1000], [1000, 3000], and [10, 1000]. Each packet contains 1080 bits. The number of GSs is 5. The transmit power of satellite and GS is 5 W and 10 W, respectively. The center carrier frequencies of satellite-to-air, ground-to-air, and ground-to-satellite links are 15 GHz, 18 GHz, and 30 GHz, respectively, each with a bandwidth of 100 MHz. ISLs employ laser bands transmitting over 197 THz. The antenna gain of satellite, aircraft and GS is 40 dB, 30 dB, and 52 dB, respectively.

\begin{figure}[t]
	\centering
	\includegraphics[width = 0.5\textwidth]{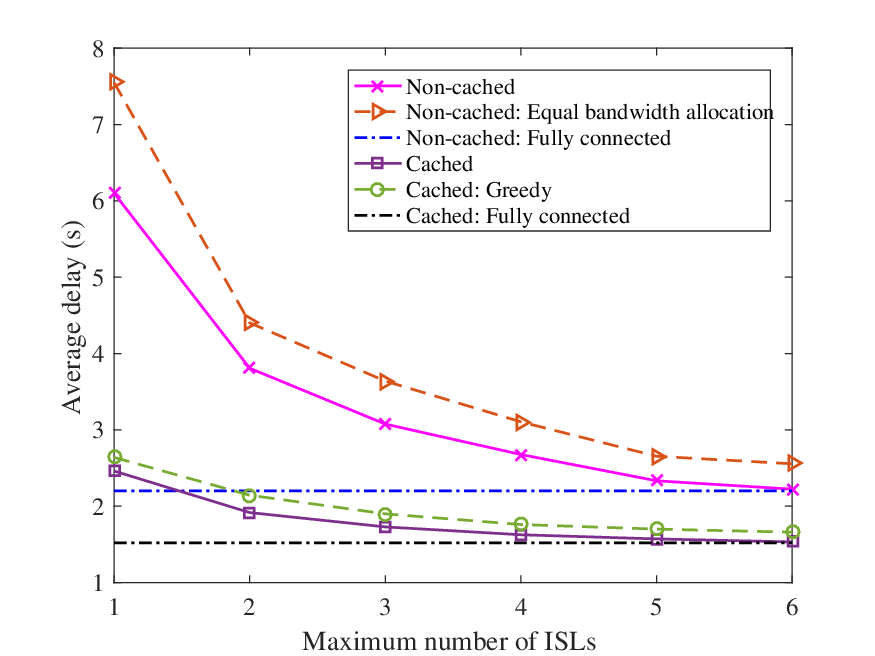}
	\caption{Average delay versus the maximum number of realizable ISLs.}\label{vs_ISL}  
\end{figure}

Fig. \ref{vs_ISL} depicts the average delay versus the maximum number of realizable ISLs.
The equal bandwidth allocation scheme for the non-cached files and greedy-based satellite association for the cached files are chosen as benchmarks. The delay when neglecting the constraint of ISLs (i.e., fully connected satellite network) is also plotted. For both cached and non-cached files, the average delay for downing the requested files decreases dramatically with the maximum number of ISLs. This is attributed to the high capacity provided by the laser ISLs, which significantly reduces the transmission delay within the satellite networks. Finally, the delay will not monotonically decrease with the number of maximum ISLs as it converges to that of the fully connected scheme. This is because the number of visible satellites that have cached the request files is limited. Therefore, the trade-off between the cost of realizing more ISLs and the performance gain it brings is still an open and important issue.

\section{Open Issues}
In this section, we identify open issues apart from the aforementioned challenges regarding ISLs.

\subsubsection{SAGIN} LEO satellite network  can be incorporated with the aerial
and ground network to constitute the space-air-ground integrated network (SAGIN), which is a dominant architecture for 6G. As mentioned above, ISLs have many application scenarios in SAGIN to ensure enhanced QoS for different tasks. Meanwhile, new standards and protocols are required to achieve a better management of the heterogeneity in SAGIN.

\subsubsection{Network Management} Future satellite network architecture is evolving to support quick configuration, seamless ISL handover, effective routing and dynamic resource allocation. These goals can be achieved by techniques such as software-defined networks (SDN) and network function virtualization (NFV), where the former decouples the control entity from the data plane to realize centralized control, while the latter separates software instances from hardware resources to implement programmable network through virtualization technologies. In 2023, Amazon launched a project named `Kuiper' that supports ISLs, where SDN is used to manage the beam scheduling and resource allocation over the constellation.

\subsubsection{RIS-Empowered ISLs} Reconfigurable intelligent surfaces (RISs) comprised of a vast number of passive elements can manipulate the propagation environment by adjusting the reflection phase of impinging signals. Besides being employed in terrestrial networks,  RISs can be mounted on LEO satellites to assist the signal propagation from one satellite to another, enabling the RIS-empowered ISLs. The use of RISs can compensate for the severe path-loss in high-frequency bands and address the high power consumption problem of a single ISL.  Several open topics for  satellite-mounted RIS communication systems should be obsessed including channel modeling and acquisition, beamforming (RIS reflection pattern) design and analysis of different performance metrics\cite{9954397}.

\subsubsection{Dedicated Techniques For LISL} Although optical ISLs have many merits like high energy concentration and low power consumption, they impose stringent requirements for pointing, acquisition and tracking (PAT). First, due to the high mobility of LEO satellites, advanced modulation and multiplexing methods need to be focused to mitigate the Doppler effects. Second, advanced space-borne laser transceivers and beam-steering schemes are required for real-time tracking of the movement and avoiding beam misalignment. Third, as the PAT processes consume much time, the link set-up time of optical ISL is longer than RF-based ISLs, which has been experimented to be tens of seconds. Hence, routing schemes for RF-based ISLs are not applicable for optical ISL as they may cause partial topology isolation\cite{10121815}.

\subsubsection{Artificial Intelligence (AI)} Owning to the efficiency and flexibility, deep learning (DL) is a prospective paradigm to optimize the performance of SAGIN in terms of network control, routing, and resource management. Since there are hundreds of deep learning architectures,  it is still tough for researchers to choose a suitable model for a specific task.  How to improve the generalization of AI-enabled techniques and their computational efficiency is a critical issue in the future.

	\section{Conclusion}\label{sec_conclusion}
ISLs can enhance LEO mega constellations by improving the network autonomy, increasing the flexibility of fulfilling various QoS requirements and providing a more favorable propagation environment. Therefore, the utilization of ISLs has received considerable attention these years. This paper gave a comprehensive overview of the vision and challenges of utilizing ISLs in LEO mega constellations. We first presented the fundamental application scenario of ISLs. Then, we revealed the main technical issues regarding ISLs from the perspectives of performance analysis, routing design and resource allocation. Finally, other research directions were outlooked to shed light on future satellite network with ISLs.

	\ifCLASSOPTIONcaptionsoff
	\newpage
	\fi

	\bibliographystyle{IEEEtran}
	\bibliography{IEEEabrv,reference.bib}

\begin{thebibliography}{10}
\providecommand{\url}[1]{#1}
\csname url@samestyle\endcsname
\providecommand{\newblock}{\relax}
\providecommand{\bibinfo}[2]{#2}
\providecommand{\BIBentrySTDinterwordspacing}{\spaceskip=0pt\relax}
\providecommand{\BIBentryALTinterwordstretchfactor}{4}
\providecommand{\BIBentryALTinterwordspacing}{\spaceskip=\fontdimen2\font plus
\BIBentryALTinterwordstretchfactor\fontdimen3\font minus \fontdimen4\font\relax}
\providecommand{\BIBforeignlanguage}[2]{{%
\expandafter\ifx\csname l@#1\endcsname\relax
\typeout{** WARNING: IEEEtran.bst: No hyphenation pattern has been}%
\typeout{** loaded for the language `#1'. Using the pattern for}%
\typeout{** the default language instead.}%
\else
\language=\csname l@#1\endcsname
\fi
#2}}
\providecommand{\BIBdecl}{\relax}
\BIBdecl

\bibitem{sagin_survey}
J.~Liu, Y.~Shi, Z.~M. Fadlullah, and N.~Kato, ``Space-air-ground integrated network: A survey,'' \emph{{IEEE} Commun. Surv. Tut.}, vol.~20, no.~4, pp. 2714--2741, Fourthquarter 2018.

\bibitem{dense}
N.~U. Hassan, C.~Huang, C.~Yuen, A.~Ahmad, and Y.~Zhang, ``Dense small satellite networks for modern terrestrial communication systems: Benefits, infrastructure, and technologies,'' \emph{{IEEE} Wireless Commun.}, vol.~27, no.~5, pp. 96--103, Oct. 2020.

\bibitem{lisl_mag}
A.~U. Chaudhry and H.~Yanikomeroglu, ``Laser intersatellite links in a starlink constellation: A classification and analysis,'' \emph{{IEEE} Veh. Technol. Mag.}, vol.~16, no.~2, pp. 48--56, Jun. 2021.

\bibitem{broadband}
Y.~Su, Y.~Liu, Y.~Zhou, J.~Yuan, H.~Cao, and J.~Shi, ``Broadband {LEO} satellite communications: Architectures and key technologies,'' \emph{{IEEE} Wireless Commun.}, vol.~26, no.~2, pp. 55--61, Apr. 2019.

\bibitem{9351765}
Q.~Chen, G.~Giambene, L.~Yang, C.~Fan, and X.~Chen, ``Analysis of inter-satellite link paths for {LEO} mega-constellation networks,'' \emph{{IEEE} Trans. Veh. Technol.}, vol.~70, no.~3, pp. 2743--2755, Mar. 2021.

\bibitem{9681176}
Q.~Chen, L.~Yang, D.~Guo, B.~Ren, J.~Guo, and X.~Chen, ``{LEO} satellite networks: When do all shortest distance paths belong to minimum hop path set?'' \emph{{IEEE} Trans. Aerosp. Electron. Syst.}, vol.~58, no.~4, pp. 3730--3734, Aug. 2022.

\bibitem{gateway}
D.~Zhou, M.~Sheng, J.~Wu, J.~Li, and Z.~Han, ``Gateway placement in integrated satellite–terrestrial networks: Supporting communications and {Internet} of remote things,'' \emph{{IEEE} Internet Things J.}, vol.~9, no.~6, pp. 4421--4434, Mar. 2022.

\bibitem{constellation_design}
Z.~Qu, G.~Zhang, H.~Cao, and J.~Xie, ``{LEO} satellite constellation for internet of things,'' \emph{IEEE Access}, vol.~5, pp. 18\,391--18\,401, 2017.

\bibitem{10121666}
N.~Zhang, Z.~Na, J.~Tao, B.~Lin, N.~Zhang, and K.~Zhao, ``Time-varying graph and binary tree search based routing algorithm for {LEO} satellite networks,'' \emph{{IEEE} Trans. Veh. Technol.}, vol.~72, no.~10, pp. 13\,683--13\,688, Oct. 2023.

\bibitem{ai}
N.~Kato, Z.~M. Fadlullah, F.~Tang, B.~Mao, S.~Tani, A.~Okamura, and J.~Liu, ``Optimizing space-air-ground integrated networks by artificial intelligence,'' \emph{{IEEE} Wireless Commun.}, Aug. 2018.

\bibitem{7805169}
X.~Jia, T.~Lv, F.~He, and H.~Huang, ``Collaborative data downloading by using inter-satellite links in {LEO} satellite networks,'' \emph{{IEEE} Trans. Wireless Commun.}, vol.~16, no.~3, pp. 1523--1532, Mar. 2017.

\bibitem{9211777}
Z.~Yan, G.~Gu, K.~Zhao, Q.~Wang, G.~Li, X.~Nie, H.~Yang, and S.~Du, ``Integer linear programming based topology design for {GNSSs} with inter-satellite links,'' \emph{{IEEE} Wireless Commun. Lett.}, vol.~10, no.~2, pp. 286--290, Feb. 2021.

\bibitem{qian}
Q.~Chen, C.~Wu, S.~Han, W.~Meng, and T.~Q. Quek, ``Exploiting inter-satellite links for in-flight connectivity scheme in space-air-ground integrated networks,'' \emph{arXiv preprint arXiv:2405.18919}, May 2024.

\bibitem{9954397}
K.~Tekbiyik, G.~K. Kurt, A.~R. Ektı, and H.~Yanikomeroglu, ``Reconfigurable intelligent surfaces empowered {THz} communication in {LEO} satellite networks,'' \emph{IEEE Access}, vol.~10, pp. 121\,957--121\,969, 2022.

\bibitem{10121815}
H.~Yang, B.~Guo, X.~Xue, X.~Deng, Y.~Zhao, X.~Cui, C.~Pang, H.~Ren, and S.~Huang, ``Interruption tolerance strategy for leo constellation with optical inter-satellite link,'' \emph{{IEEE} Trans. Netw. Service Manag.}, vol.~20, no.~4, pp. 4815--4830, Dec. 2023.

\end{thebibliography}

\end{document}